# Modulated Ringdown Comb Interferometry for next-generation high complexity trace gas sensing

Qizhong Liang[1,*], Apoorva Bisht[1], Andrew Scheck[1], Peter G. Schunemann[2,#], and Jun Ye[1,*]

[1] JILA, National Institute of Standards and Technology and University of Colorado, Boulder, CO 80309

[2] BAE Systems, Nashua, NH 03061

[#] Present address: Onsemi, Hudson, NH 03051

*Corresponding authors: Qizhong.Liang@colorado.edu, Ye@jila.colorado.edu

Gas samples relevant to health[1-3] and environment[4-6] typically contain a plethora of molecular species that span a huge concentration dynamic range. High-concentration molecules impose a strong absorption background that hinders robust identification of low-concentration species. While mid-infrared frequency comb spectroscopy with high-finesse cavity enhancement has realized many of the most sensitive multi-species trace gas detection to date[2,7-13], its robust performance requires gas samples to contain only weak absorption features to avoid dispersing cavity resonances from the comb line frequencies. Here we introduce a new technique that is free from this restriction, thus enabling the development of next-generation multi-species trace gas sensing with broad applicability to complex and dynamic molecular compositions. The principle of Modulated Ringdown Comb Interferometry is to resolve ringdown dynamics carried by massively parallel comb lines transmitted through a length-modulated cavity. This method leverages both periodicity of the field dynamics and Doppler frequency shifts introduced from a Michelson interferometer. Scalable enhancement of both spectral coverage and cavity finesse is enabled with dispersion immune and high-efficiency data collection. Built upon this platform, we realize in the mid-infrared a product of finesse and spectral coverage that is orders of magnitude better than all prior experiments[2,7-19]. We demonstrate the power of this technique by measuring highly dispersive exhaled human breath samples over a vastly expanded spectral coverage of 1,010 cm$^{-1}$ and with cavity finesse of 23,000. This allows for the first time simultaneous quantification of 20 distinct molecular species at > 1 part-per-trillion sensitivity with their concentrations varying by 7 orders of magnitude.

**Introduction:**

Achieving the most sensitive multi-species trace gas sensing is critical for many frontier applications including precise determination of complex molecular structures[9,11], real-time reaction kinetics[12,13], atmospheric sensing[4-6], and breath-based medical diagnostics[1-3]. Orders of magnitude enhancement in the light-matter interaction length can be achieved over a broad spectral coverage by having an optical frequency comb coupled into a high-finesse cavity surrounding a gas sample. However, to date, this method is intrinsically vulnerable to intracavity dispersion: the stronger the absorption signals, the larger the distortion of cavity resonances from the comb line frequencies to restrict the spectral acceptance. This compromises the use of higher cavity finesse for stronger absorption signals with the desire to also achieve wider spectral coverage for a more diverse set of molecular bonds.

For breath-based medical diagnostics, improved diagnostic accuracy will result from the detection of a larger number of molecular species that are discriminative in concentrations between medical conditions[3]. The desire to accurately detect complex mixtures of trace molecules of clinical relevance in their native environment with far better than one part-per-billion sensitivity must be carried out in the presence of high concentrations of $H_2O$ and $CO_2$. In our recently established sensing performance for exhaled breath[2], we are limited to a single-shot comb-cavity coupled bandwidth of < 10 $cm^{-1}$ to avoid significant frequency-to-amplitude noise conversion due to dispersion from breath molecules. A total of 135 $cm^{-1}$ coverage (2,810-2,945 $cm^{-1}$) with spectral concatenation and sensitivity at the level of 100 parts-per-trillion was achieved. Further spectral extension beyond the current range is highly challenging due to the intense absorption lines from $H_2O$, HDO, and $CH_4$. This has restricted us to the detection of primarily hydrocarbon species, placing a limit on the diagnostics utility and performance[3].

Here, we introduce a novel spectroscopy technique based on Modulated Ringdown Comb Interferometry (MRCI) to make possible the most sensitive multi-species trace gas sensing detection regardless of whether the gas samples contain strongly absorbing compounds. We employ a scanning comb-cavity coupling scheme for dispersion immunity while at the same time realizing high-resolution cavity ringdown readout to ensure extremely sensitive and broadband trace gas detection. We perform highly efficient data collection with a single-shot coverage at the full instantaneous comb bandwidth ~ 200 $cm^{-1}$ (x 20 improvement) and a total spectral coverage of 1,010 $cm^{-1}$ after concatenation. This is a new capability to navigate through all the intense absorption features from molecules including $H_2O$, HDO, $CH_4$, $CO_2$, and $N_2O$ to detect the lowest concentration species at sensitivity of a few parts-per-trillion (x 100 improvement). The basic concept shown in Fig. 1 is to repeatedly sweep each cavity resonance through a corresponding comb line at a modulation frequency $\omega_m$, thus ensuring every comb line gets coupled into the cavity. The cavity transmission is monitored with a Michelson interferometer passively scanning at a constant speed $v$. In the time domain, the periodic cavity transmission of each comb line is modulated by the interferometer readout at its corresponding Doppler frequency $\omega_i \propto k_i v$, where $k_i$ is the wavenumber of the $i$-th comb line. Only two conditions need to be met: first, the acquisition time per interferogram $t$ must be longer than $2\pi/\Delta\omega_i$ for comb-resolved resolution, where $\Delta\omega_i = \omega_{i+1} - \omega_i$; second, the interferometer scanning speed $v$ should be chosen such that the product of $\Delta\omega_i$ and the total number of comb lines is smaller than $\omega_m$. In doing so, Fourier-transform of the interferogram yields multiple sets of RF combs, with the set generated by the $i$-th comb line exhibits a frequency spacing of $\omega_m$ and offset of $\omega_i$, as shown in Fig. 1c. Massively parallel measurement of ringdown value at each comb line frequency is thus realized by analyzing the spectral intensity distribution for each set of RF comb.

This approach also comes with the advantage of simple implementation with a free-running comb requiring no high bandwidth servos and complicated synchronization schemes. As a result, this technique unleashes a powerful way to address many of the challenging real-world trace gas sensing tasks[20-23] by providing unmatched detection sensitivity, wide applicability to gas samples of complex and dynamic composition, and improved practicality and robustness for deployment. We report our new apparatus at much improved finesse-spectral coverage product (Fig. 2) to showcase this newly established state-of-the-art performance demonstrated over highly dispersive exhaled breath samples.

**Principle of Modulated Ringdown Comb Interferometry (MRCI):**

The spectroscopy technique of MRCI encompasses two essential ingredients (Fig. 1a): i) a high-finesse cavity swept-locked to the incident mid-infrared comb to permit its periodic transmission through the cavity once per cavity length sweep. ii) An asynchronous, passively-scanning Michelson interferometer to read out the periodic cavity transmission bursts through balanced detection.

Divided in steps (Fig. 1b), we present how the recorded interferogram as a function of time allows high-spectral resolution determination of the ringdown time spectrum. Step 1) Consider a single transmission intensity burst generated from the $i$-th cavity resonance. Related by a Fourier transformation, the shape of the spectral envelope uniquely determines the ringdown time. A shorter ringdown time corresponds to a flatter spectral envelope. Step 2) Record the same transmission bursts repeatedly at a periodicity $2\pi/\omega_m$: The original spectral information is now transformed into a set of Fourier components spaced by $\omega_m$. Step 3) Record the same periodic transmission bursts through balanced detection from a Michelson interferometer. All Fourier components are frequency displaced with a corresponding Doppler shift that is proportional to the $i$-th cavity resonance wavenumber and velocity of the delay stage. The spectral information is thus engineered into an "RF spectral comb", with a spacing of $\omega_m$ and spectral intensity distribution determined by the cavity ringdown time. Multiple "RF spectral combs" generated from optical comb components transmitted through different cavity resonances share the same spacing $\omega_m$ but with discrete frequency offsets (Fig. 1c). Determination of the spectral intensity distributions for multiple sets of "RF spectral combs" thus realizes broadband simultaneous ringdown measurements.

The dispersion immunity brought by the MRCI technique allows scalable expansion of spectral coverage simply by adding the number of standard high-finesse cavities coated at different wavelength ranges. In our newly constructed apparatus, we employed two high-finesse cavities with axes intersected to probe the same gas sample (Fig. 3a). One cavity permits spectroscopic data collection over the 1,850-2,230 cm$^{-1}$ wavenumber range (near 5-μm in wavelength; finesse peaks at 23,000; ringdown time peaks at 17 μs), the other at the 2,700-3,330 cm$^{-1}$ range (near 3-μm in wavelength; finesse peaks at 14,000; ringdown time peaks at 8 μs). Data collection for either cavity is highly automated (Methods). The comb source used for the 5-μm cavity is a newly constructed synchronously pumped, singly resonant optical parametric oscillator using a type-I phase-matched zinc germanium phosphide crystal (Methods). The 3-μm cavity is probed with a previously reported OPO comb[24].

**Demonstration of MRCI over highly dispersive complex gas samples:**

We demonstrate the powerful capability of MRCI with measurements of exhaled breath collected from nasal and oral respiratory airways and ambient air (see Methods for sampling protocol). The new record-level finesse and broadband coverage allow us to accurately quantify subtle differences in these complex gas samples for a comprehensive list of trace molecular species. Notably, nitric oxide, the species with administrative approval for asthma monitoring, can now be robustly measured on top of the strongly saturated water absorption background and down to its extremely low concentration of ~20 ppb[25].

Sample data collected with 2 seconds of acquisition time with the cavity filled with an exhaled breath sample using MRCI are presented in Fig. 3b. Trend lines in red highlight the two sets of Fourier components generated at two different optical frequencies. Insets show the fitted ringdown values. A faster drop in spectral intensities with the increase of Fourier harmonics order corresponds to a longer cavity ringdown time. Measured for the same breath sample, full-coverage ringdown spectra determined from the 5-µm cavity is presented in Fig. 3c, and from the 3-µm cavity is presented in Fig. 3d. Comparing the ringdown data measured with cavity loaded with breath against that measured with empty cavity (held at base pressure below 3 mTorr), the reduction in the ringdown time as a function of optical frequency yields a plethora of molecular absorption features generated uniquely from the loaded breath molecules inside the cavity. Ringdown data measured for the empty 5-µm cavity correspond to water absorption signals at below 2,000 $cm^{-1}$, arising from residual water molecules present inside the cavity. The ringdown spectrum measured for the empty 3-µm cavity was found to exhibit large but slowly varying reflectivity oscillating at a period of about 300 $cm^{-1}$ wavenumbers, a feature likely due to the mirror coating design. For both cavities, intracavity dispersion from neither the non-flat mirror spectral response nor the strong molecular absorptions prohibit us from utilizing the entire range of high-reflectivity coating for ultra-sensitive broadband absorption spectroscopy.

Ringdown data collected for various gas samples are referenced against the empty cavity data for the determination of the molecular absorption spectra. Data collected from the 5- and 3-µm cavities are presented respectively in Figure 4a and Figure 5a**.** Global molecular fitting results to the HITRAN database[26] are plotted as inverted features. Zoom-ins to discrete spectral regions, where individual molecular fit results are plotted with color to highlight absorption features from nitric oxide (NO), carbon monoxide isotopologue ($^{13}CO$), formaldehyde ($H_2CO$), and methanol ($CH_3OH$), as shown respectively in Fig. 4b, 4c, 5b, and 5c. For nitric oxide and formaldehyde, their absorption features are much stronger in nasal breath samples, but weaker and similar in both oral breath and ambient air. $^{13}CO$ and methanol's absorption features are visually less contrasting between nasal and oral breaths but considerably more intense than that from the air sample. For these extremely low-concentration molecular species, it can be clearly seen that their weak absorption features are highly overlapped with absorption features from other strongly absorbing species, yet they are robustly measured by MRCI with dispersion immunity.

Molecular cross-section data totaling 20 species is fitted to the experimental data. Summarized in Fig. 6, we identified strongly elevated concentrations for nitric oxide, formaldehyde, acetone ($CH_3COCH_3$), water isotopologues ($H_2O$, $H_2^{18}O$, $H_2^{17}O$, and HDO) in nasal breath compared with oral breath. We attribute the results for nitric oxide[25,27] and carbon monoxide isotopologues (CO, $^{13}CO$)[28] to the production in the paranasal sinuses, while for water isotopologue to the supply from the mucous membrane. Elevations in formaldehyde and acetone are plausibly due to their high water solubility and thus carried out more by the increased water contents in the nasal airway. On the other hand, elevation in nitrous oxide ($N_2O$) in the oral breath might arise from production by denitrifying bacteria present in the oral cavity[29]. We note that molecular profiles measured from nasal and oral breaths are likely to vary from one research subject to the next and subject to change depending on the sampling methods and their actual execution. The current study focuses primarily on reporting the new capabilities for characterizing highly dispersive gas samples. Nevertheless, this powerful capability has allowed us to establish a broader utility to observe simultaneously a rich variety of biological conditions through non-invasive measurements of breath.

Comparing our breath measurement results with that of air, carbon dioxide isotopologues ($CO_2$, $^{13}CO_2$, $OC^{18}O$, $OC^{17}O$), water isotopologues ($H_2O$, $H_2^{18}O$, $H_2^{17}O$, and HDO), methane isotopologues ($CH_4$, $^{13}CH_4$) and carbon monoxide isotopologues (CO, $^{13}CO$) measured from the breath samples are strongly elevated from that in air. Strong absorption features from these species explains why exhaled breath is much more dispersive than ambient air, hence much more challenging to measure if employing tight comb-cavity coupling schemes. Elevation in methane confirms the research subject as a methane producer, a result of excess archaea microbiome presented in the small intestine[30]. The typical concentrations of the fitted molecular species measured for healthy controls reported by various literature are listed in circles on the stem lines in Figure 6. They generally agree well with our measurement results. While most literature is limited to the measurement of only one or a few molecule species, here our experimental capability allows us to examine all 20 species at once. Noise equivalent detection sensitivity was analyzed using the spectroscopy data collected from the cavity held at base pressure following the same analysis protocol detailed in Ref.[2]. The minimum detectable absorption per spectral element at one second is $3 \times 10^{-10}$ cm$^{-1}$Hz$^{-1/2}$ for the 5-μm cavity and $6 \times 10^{-10}$ cm$^{-1}$Hz$^{-1/2}$ for the 3-μm cavity. Normalized to the acquisition time of 100 s, molecular concentration sensitivities into the parts-per-trillion levels (below 1 ppb) are found for 15 out of the 20 molecules and are summarized by the circles on the stem lines in Figure 6. The best sensitivity at 8 ppt is obtained for N$^{15}$NO and $^{15}$NNO. More details on data fitting can be found in Methods.

**Discussion:**

For practical applications of comb-based trace gas detection, two highly desired features immediately come to mind: an outstanding versatility to detect both narrow and broad, or strong and weak spectral features, and a simple and robust operation for field applications. We now discuss these two features. The MRCI technique presented here enables molecular absorption spectra to be robustly determined against a broad background and referenced directly to a constant baseline determined by the reflectivity of the cavity mirrors. This provides an exceptional versatility to the technique for demanding applications of measuring absorption features of arbitrary spectral widths beyond that of the instantaneous bandwidth of the laser source[31]. As evident from the presented results, ro-vibrational quantum state-resolved absorption signals from small molecules with a characteristic spectral width of ~0.1-1.0 GHz (e.g., nitric oxide) and state-unresolved signals from medium-size molecules at ~10-100 cm$^{-1}$ (e.g., acetone and methanol) are all robustly measured. This unique and outstanding capability paves the way to detect viruses[32,33] and proteins[34] that are potentially present in breath, where absorption features span over 1,000 cm$^{-1}$. Direct quantification of viral loads in exhaled breath can allow differentiation of viral response from host response in diagnostic scenarios such as detection of SARS-CoV-2 infection to yield extra mechanistic insights[3].

We now turn to the practicality of various spectroscopy methods. Traditionally, measurement of a cavity ringdown time is carried out by spatially dispersing the cavity transmission fields followed by fitting the spectral intensity data measured as a function of time[35]. Such a detection scheme is not practical in the mid-infrared spectroscopic region due to the slow integration time (≥ 10 μs) of detector arrays to resolve ringdown dynamics[12-14]. Instead, cavity-transmitted intensity measurements rather than ringdown are typically performed. Highly dispersive virtually imaged phase arrays etalons are required to achieve sufficient spectral resolution (< 1 GHz) and the instantaneous spectral coverage is compromised to well below 100 cm$^{-1}$, limited by the size of the detector arrays. High optical power loss and strong detection noise floor impose further technological restrictions to achieve the highest possible detection sensitivity.

Here, spectroscopy readout by a Michelson interferometer simultaneously allows spectroscopic measurements down to comb-mode resolution, high electronic detection bandwidth for ringdown measurements, and high optical power utilization efficiency. In contrast to dual-comb readout[36-39], Michelson interferometer removes the need to construct a second comb source of tight mutual coherence with the first comb, resulting in a lower implementation barrier, smaller setup cost, and more robust operation. MRCI adopts the same strategy as dual-comb spectroscopy to position spectral information carriers uniformly over the entire electronic detection bandwidth to achieve fast data readout time.

All frequency comb sources used in this experiment were constructed using mature technologies from fiber mode-locked oscillators and optical parametric oscillators. Replacing tight comb-cavity coupling with a cavity-swept lock enables dramatic improvement to data collection efficiency: it permits the entire instantaneous comb bandwidth to couple through the cavity despite presence of strong intracavity dispersion, and require no comb frequencies to be measured and locked to cavity resonances with high servo bandwidth[16,40-43]. A cavity swept lock also enables a large dynamic range for servo and robust sustentation against incidental mechanical jittering, making it much more practical for out-of-laboratory operation. Compatible with combs under free-running condition, MRCI can be implemented with comb sources produced with integrated photonics[18,44,45] to facilitate the construction of highly compact and portable devices. The cavity ringdown detection principle also provides immunity from laser intensity noise, optical etalons, and molecular absorptions external to the cavity. It is also free from measurement calibration and allows robust and reliable operation regardless of long-term variations in laser conditions and optical alignment drifts. Routine maintenance is thus minimized while ensuring spectroscopic data are acquired with much-improved accuracy and reliability.

**Conclusion:**

We present a novel spectroscopy technique to equip cavity-enhanced frequency comb spectroscopy with dispersion-immune, massively parallel ringdown measurement capability. This new platform not only enables the next-generation highly versatile sensing tools to be developed, but also a much simpler technological implementation: 1) Weakest, strongest, sharpest, and broadest absorption features are simultaneously and robustly measured; 2) Future upgrade from the current state-of-the-art is now feasible with scalable enhancement in spectral coverage and cavity finesse; 3) Comb sources can be operated under free-running condition and no high-bandwidth feedback servos are required; 4) Robust, automatable and high-efficiency data collection is now feasible requiring little to no trained expertise. Lastly, the modulated ringdown comb interferometry technique should be conceived as a massively multiplexed information extraction methodology, where channels of information encoded into the field intensity dynamics of frequency comb lines can be determined in parallel from its frequency domain equivalent carriers. Extension of this measurement methodology to beyond just the ringdown time should be straightforward. For example, to measure simultaneously the reaction kinetics for multiple short-lived chemical species generated at fixed time periodicity[12], where real-time absorptions encoded to comb lines can be parallelly measured up to the response bandwidth of photodetectors.

**References:**

1. Thorpe, M. J., Balslev-Clausen, D., Kirchner, M. S. & Ye, J. Cavity-enhanced optical frequency comb spectroscopy: application to human breath analysis. *Opt Express* **16**, 2387-2397 (2008). doi:10.1364/Oe.16.002387
2. Liang, Q. *et al.* Ultrasensitive multispecies spectroscopic breath analysis for real-time health monitoring and diagnostics. *Proc Natl Acad Sci* **118** (2021). doi:10.1073/pnas.2105063118
3. Liang, Q. *et al.* Breath analysis by ultra-sensitive broadband laser spectroscopy detects SARS-CoV-2 infection. *J Breath Res* **17** (2023). doi:10.1088/1752-7163/acc6e4
4. Rieker, G. B. *et al.* Frequency-comb-based remote sensing of greenhouse gases over kilometer air paths. *Optica* **1**, 290-298 (2014). doi:10.1364/Optica.1.000290
5. Herman, D. I. *et al.* Precise multispecies agricultural gas flux determined using broadband open-path dual-comb spectroscopy. *Science Advances* **7** (2021). doi:10.1126/sciadv.abe9765
6. Giorgetta, F. R. *et al.* Open-Path Dual-Comb Spectroscopy for Multispecies Trace Gas Detection in the 4.5-5 µm Spectral Region. *Laser Photonics Rev* **15** (2021). doi:10.1002/lpor.202000583
7. Bui, T. Q. *et al.* Spectral analyses of trans- and cis-DOCO transients via comb spectroscopy. *Mol Phys* **116**, 3710-3717 (2018). doi:10.1080/00268976.2018.1484949
8. Changala, P. B., Spaun, B., Patterson, D., Doyle, J. M. & Ye, J. Sensitivity and resolution in frequency comb spectroscopy of buffer gas cooled polyatomic molecules. *Appl Phys B* **122** (2016). doi:10.1007/s00340-016-6569-7
9. Spaun, B. *et al.* Continuous probing of cold complex molecules with infrared frequency comb spectroscopy. *Nature* **533**, 517-520 (2016). doi:10.1038/nature17440
10. Foltynowicz, A., Maslowski, P., Fleisher, A. J., Bjork, B. J. & Ye, J. Cavity-enhanced optical frequency comb spectroscopy in the mid-infrared application to trace detection of hydrogen peroxide. *Appl Phys B* **110**, 163-175 (2013). doi:10.1007/s00340-012-5024-7
11. Changala, P. B., Weichman, M. L., Lee, K. F., Fermann, M. E. & Ye, J. Rovibrational quantum state resolution of the C60 fullerene. *Science* **363**, 49-54 (2019). doi:10.1126/science.aav2616
12. Bjork, B. J. *et al.* Direct frequency comb measurement of OD + CO --> DOCO kinetics. *Science* **354**, 444-448 (2016). doi:10.1126/science.aag1862
13. Bui, T. Q. *et al.* Direct measurements of DOCO isomers in the kinetics of OD + CO. *Science Advances* **4** (2018). doi:10.1126/sciadv.aao4777
14. Fleisher, A. J. *et al.* Mid-Infrared Time-Resolved Frequency Comb Spectroscopy of Transient Free Radicals. *J Phys Chem Lett* **5**, 2241-2246 (2014). doi:10.1021/jz5008559
15. Lu, C., Morville, J., Rutkowski, L., Vieira, F. S. & Foltynowicz, A. Cavity-Enhanced Frequency Comb Vernier Spectroscopy. *Photonics-Basel* **9** (2022). doi:10.3390/photonics9040222
16. Sulzer, P. *et al.* Cavity-enhanced field-resolved spectroscopy. *Nat Photonics* **16**, 692-+ (2022). doi:10.1038/s41566-022-01057-0
17. Khodabakhsh, A. *et al.* Fourier transform and Vernier spectroscopy using an optical frequency comb at 3-5.4 µm. *Opt Lett* **41**, 2541-2544 (2016). doi:10.1364/Ol.41.002541
18. Sterczewski, L. A. *et al.* Cavity-Enhanced Vernier Spectroscopy with a Chip-Scale Mid-Infrared Frequency Comb. *Acs Photonics* **9**, 994-1001 (2022). doi:10.1021/acsphotonics.1c01849
19. Haakestad, M. W., Lamour, T. P., Leindecker, N., Marandi, A. & Vodopyanov, K. L. Intracavity trace molecular detection with a broadband mid-IR frequency comb source. *J Opt Soc Am B* **30**, 631-640 (2013). doi:10.1364/Josab.30.000631
20. Costello, B. D. *et al.* A review of the volatiles from the healthy human body. *Journal of Breath Research* **8** (2014). doi:10.1088/1752-7155/8/1/014001
21. Deshmukh, C. S. *et al.* Net greenhouse gas balance of fibre wood plantation on peat in Indonesia. *Nature* **616**, 740-+ (2023). doi:10.1038/s41586-023-05860-9

22	van Groenigen, K. J., Osenberg, C. W. & Hungate, B. A. Increased soil emissions of potent greenhouse gases under increased atmospheric CO2. *Nature* **475**, 214-U121 (2011). doi:10.1038/nature10176
23	Guenther, A. *et al.* Natural emissions of non-methane volatile organic compounds; carbon monoxide, and oxides of nitrogen from North America. *Atmos Environ* **34**, 2205-2230 (2000). doi:10.1016/S1352-2310(99)00465-3
24	Adler, F. *et al.* Phase-stabilized, 1.5 W frequency comb at 2.8-4.8 μm. *Opt Lett* **34**, 1330-1332 (2009). doi:10.1364/Ol.34.001330
25	Dweik, R. A. *et al.* An Official ATS Clinical Practice Guideline: Interpretation of Exhaled Nitric Oxide Levels (FENO) for Clinical Applications. *Am J Resp Crit Care* **184**, 602-615 (2011). doi:10.1164/rccm.9120-11ST
26	Gordon, I. E. *et al.* The HITRAN2020 molecular spectroscopic database. *J Quant Spectrosc Ra* **277** (2022). doi:10.1016/j.jqsrt.2021.107949
27	Lundberg, J. O. N. *et al.* High Nitric-Oxide Production in Human Paranasal Sinuses. *Nat Med* **1**, 370-373 (1995). doi:10.1038/nm0495-370
28	Andersson, J. A., Uddman, R. & Cardell, L. O. Carbon monoxide is endogenously produced in the human nose and paranasal sinuses. *J Allergy Clin Immun* **105**, 269-273 (2000). doi:10.1016/S0091-6749(00)90075-7
29	Mitsui, T., Miyamura, M., Matsunami, A., Kitagawa, K. & Arai, N. Measuring nitrous oxide in exhaled air by gas chromatography and infrared photoacoustic spectrometry. *Clin Chem* **43**, 1993-1995 (1997). doi:10.1093/clinchem/43.10.1993
30	Costello, B. P. J. D., Ledochowski, M. & Ratcliffe, N. M. The importance of methane breath testing: a review. *Journal of Breath Research* **7** (2013). doi:10.1088/1752-7155/7/2/024001
31	Pupeza, I. *et al.* Field-resolved infrared spectroscopy of biological systems. *Nature* **577**, 52-+ (2020). doi:10.1038/s41586-019-1850-7
32	Xia, Q. *et al.* Single virus fingerprinting by widefield interferometric defocus-enhanced mid-infrared photothermal microscopy. *Nat Commun* **14** (2023). doi:10.1038/s41467-023-42439-4
33	Barauna, V. G. *et al.* Ultrarapid On-Site Detection of SARS-CoV-2 Infection Using Simple ATR-FTIR Spectroscopy and an Analysis Algorithm: High Sensitivity and Specificity. *Anal Chem* **93**, 2950-2958 (2021). doi:10.1021/acs.analchem.0c04608
34	López-Lorente, A. I. & Mizaikoff, B. Mid-infrared spectroscopy for protein analysis: potential and challenges. *Anal Bioanal Chem* **408**, 2875-2889 (2016). doi:10.1007/s00216-016-9375-5
35	Thorpe, M. J., Moll, K. D., Jones, R. J., Safdi, B. & Ye, J. Broadband cavity ringdown spectroscopy for sensitive and rapid molecular detection. *Science* **311**, 1595-1599 (2006). doi:10.1126/science.1123921
36	Muraviev, A. V., Smolski, V. O., Loparo, Z. E. & Vodopyanov, K. L. Massively parallel sensing of trace molecules and their isotopologues with broadband subharmonic mid-infrared frequency combs. *Nat Photonics* **12**, 209-+ (2018). doi:10.1038/s41566-018-0135-2
37	Ycas, G. *et al.* High-coherence mid-infrared dual-comb spectroscopy spanning 2.6 to 5.2 μm. *Nat Photonics* **12**, 202-+ (2018). doi:10.1038/s41566-018-0114-7
38	Long, D. A. *et al.* Nanosecond time-resolved dual-comb absorption spectroscopy. *Nat Photonics* **18**, 127-131 (2024). doi:10.18434/mds2-3091
39	Pupeikis, J. *et al.* Spatially multiplexed single-cavity dual-comb laser. *Optica* **9**, 713-716 (2022). doi:10.1364/Optica.457787
40	Bernhardt, B. *et al.* Cavity-enhanced dual-comb spectroscopy. *Nat Photonics* **4**, 55-57 (2010). doi:10.1038/Nphoton.2009.217
41	Dubroeucq, R. & Rutkowski, L. Optical frequency comb Fourier transform cavity ring-down spectroscopy. *Opt Express* **30**, 13594-13602 (2022). doi:10.1364/Oe.454775
42	Lisak, D. *et al.* Dual-comb cavity ring-down spectroscopy. *Sci Rep-Uk* **12** (2022). doi:10.1038/s41598-022-05926-0

43	Hoghooghi, N. *et al.* Broadband coherent cavity-enhanced dual-comb spectroscopy. *Optica* **6**, 28-33 (2019). doi:10.1364/Optica.6.000028
44	Suh, M. G., Yang, Q. F., Yang, K. Y., Yi, X. & Vahala, K. J. Microresonator soliton dual-comb spectroscopy. *Science* **354**, 600-603 (2016). doi:10.1126/science.aah6516
45	Stokowski, H. S. *et al.* Integrated frequency-modulated optical parametric oscillator. *Nature* **627** (2024). doi:10.1038/s41586-024-07071-2
46	Tian, L. *et al.* Gas phase multicomponent detection and analysis combining broadband dual-frequency comb absorption spectroscopy and deep learning. *Communications Engineering* **2**, 54 (2023). doi:10.1038/s44172-023-00105-z
47	Zhu, F. *et al.* Mid-infrared dual frequency comb spectroscopy based on fiber lasers for the detection of methane in ambient air. *Laser Phys Lett* **12** (2015). doi:10.1088/1612-2011/12/9/095701
48	Johnson, T. A. & Diddams, S. A. Mid-infrared upconversion spectroscopy based on a Yb:fiber femtosecond laser. *Appl Phys B* **107**, 31-39 (2012). doi:10.1007/s00340-011-4748-0
49	Hjältén, A., Foltynowicz, A. & Sadiek, I. Line positions and intensities of the v1 band of 12CH3I using mid-infrared optical frequency comb Fourier transform spectroscopy. *J Quant Spectrosc Ra* **306** (2023). doi:10.1016/j.jqsrt.2023.108646
50	Zuo, Z. *et al.* Broadband mid-infrared molecular spectroscopy based on passive coherent optical-optical modulated frequency combs. *Photonics Res* **9**, 1358-1368 (2021). doi:10.1364/Prj.422397
51	Tomaszewska-Rolla, D. *et al.* Mid-infrared optical frequency comb spectroscopy using an all-silica antiresonant hollow-core fiber. *Opt Express* **32**, 10679-10689 (2024). doi:10.1364/OE.517012
52	Adler, F. *et al.* Mid-infrared Fourier transform spectroscopy with a broadband frequency comb. *Opt Express* **18**, 21861-21872 (2010). doi:10.1364/Oe.18.021861
53	Sterczewski, L. A. *et al.* Mid-infrared dual-comb spectroscopy with interband cascade lasers. *Opt Lett* **44**, 2113-2116 (2019). doi:10.1364/Ol.44.002113
54	Komagata, K. N., Wittwer, V. J., Südmeyer, T., Emmenegger, L. & Gianella, M. Absolute frequency referencing for swept dual-comb spectroscopy with midinfrared quantum cascade lasers. *Phys Rev Res* **5** (2023). doi:10.1103/PhysRevResearch.5.013047
55	Hjältén, A. *et al.* Optical frequency comb Fourier transform spectroscopy of 14N216O at 7.8 μm. *J Quant Spectrosc Ra* **271** (2021). doi:10.1016/j.jqsrt.2021.107734
56	Germann, M. *et al.* A methane line list with sub-MHz accuracy in the 1250 to 1380 cm-1 range from optical frequency comb Fourier transform spectroscopy. *J Quant Spectrosc Ra* **288** (2022). doi:10.1016/j.jqsrt.2022.108252
57	Nielsen, G. D. & Wolkoff, P. Cancer effects of formaldehyde: a proposal for an indoor air guideline value. *Arch Toxicol* **84**, 423-446 (2010). doi:10.1007/s00204-010-0549-1
58	Wang, Z. N. & Wang, C. J. Is breath acetone a biomarker of diabetes? A historical review on breath acetone measurements. *Journal of Breath Research* **7** (2013). doi:10.1088/1752-7155/7/3/037109
59	Turner, C., Spanel, P. & Smith, D. A longitudinal study of methanol in the exhaled breath of 30 healthy volunteers using selected ion flow tube mass spectrometry, SIFT-MS. *Physiol Meas* **27**, 637-648 (2006). doi:10.1088/0967-3334/27/7/007
60	Dryahina, K., Smith, D. & Spanel, P. Quantification of methane in humid air and exhaled breath using selected ion flow tube mass spectrometry. *Rapid Commun Mass Sp* **24**, 1296-1304 (2010). doi:10.1002/rcm.4513
61	Paredi, P., Kharitonov, S. A. & Barnes, P. J. Elevation of exhaled ethane concentration in asthma. *Am J Resp Crit Care* **162**, 1450-1454 (2000). doi:10.1164/ajrccm.162.4.2003064
62	Cunnington, A. J. & Hormbrey, P. Breath analysis to detect recent exposure to carbon monoxide. *Postgrad Med J* **78**, 233-237 (2002). doi:10.1136/pmj.78.918.233
63	Mansour, E. *et al.* Measurement of temperature and relative humidity in exhaled breath. *Sensor Actuat B-Chem* **304** (2020). doi:10.1016/j.snb.2019.127371

64      Dawson, B. *et al.* Measurements of methane and nitrous oxide in human breath and the development of UK scale emissions. *Plos One* **18** (2023). doi:10.1371/journal.pone.0295157

**Acknowledgements:**

We thank Gregory B. Rieker, Scott A. Diddams, Thinh Q. Bui, Oliver H. Heckl, Adam J. Fleisher, P. Bryan Changala, and David J. Nesbitt for helpful discussions. This work is supported by AFOSR, NIST, NSF QLCI OMA–2016244 and NSF PHY-1734006.

**Author contributions**

All authors contributed to the manuscript writing and results interpretation. Q. L. and J. Y. conceived the modulated ringdown comb interferometry technique. Q. L., A. B., and A. S. designed and constructed the spectroscopy setup and collected and analyzed the spectroscopy data. The 5-μm optical parametric oscillator comb was designed and constructed by Q. L. using the zinc germanium phosphide nonlinear optical crystal provided by P. G. S. The research work was supervised by J. Y.

**Competing interests:**

The authors declare no competing interests.

**Data availability:**

All data supporting the findings of this study are available within the paper.

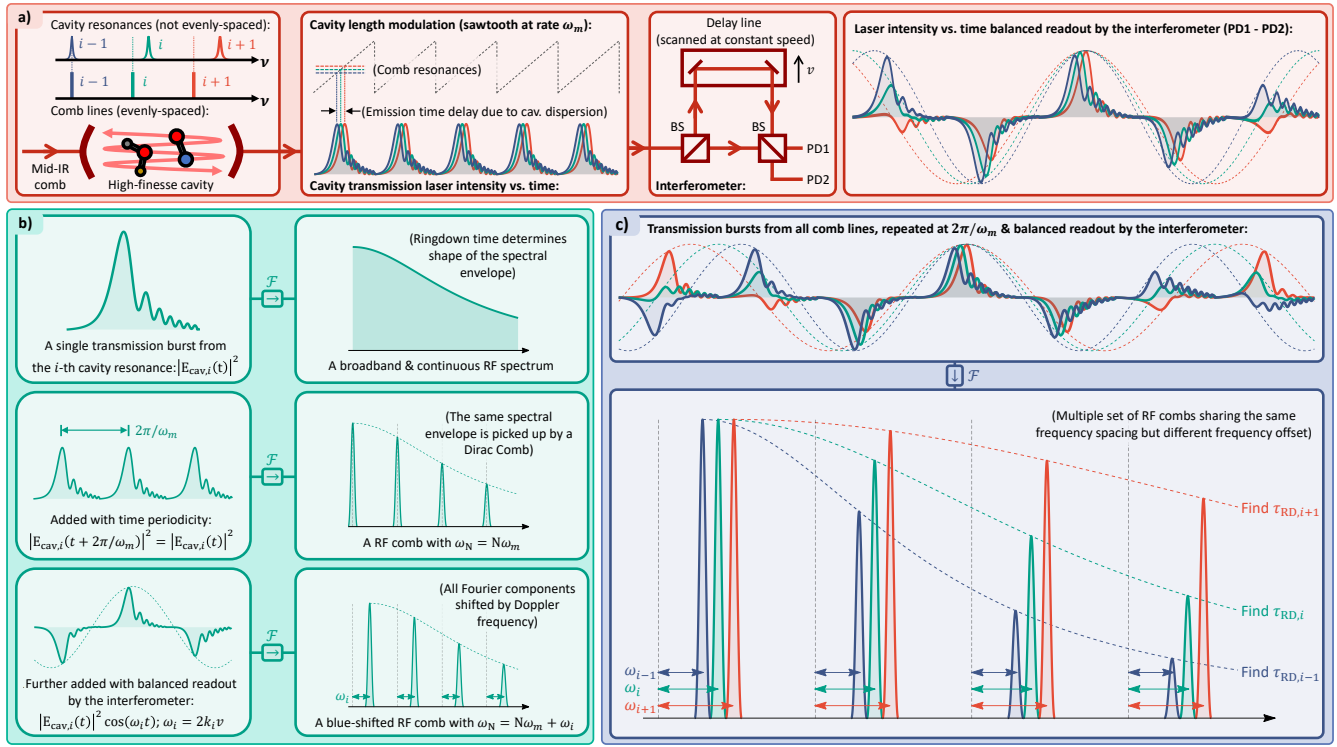

**Figure 1: The Modulated Ringdown Comb Interferometry (MRCI) technique. a,** A high-finesse cavity periodically modulated in cavity length, permits incident mid-infrared comb to transmit through the cavity once per cavity length sweep. Within a single sweep, laser fields from different cavity resonances are emitted in sequence as a result of intracavity dispersion determined by the loaded gas molecules and mirror spectral response. The periodic cavity transmission bursts are readout by a passively-scanning Michelson interferometer via balanced detection to yield an interferogram measured for laser intensity as a function of time. **b,** Physical interpretations to the recorded interferogram. Shown are the intensity subcomponents vs. time generated from a particular cavity resonance. Through addition of time periodicity and readout by interferometry, the spectral information generated by the same cavity resonance is engineered into an "RF spectral comb" with frequency spacing given by cavity length modulation frequency, frequency offset given by the Doppler frequency, and a spectral envelope determined by the ringdown time. **c,** Fourier transform of the recorded interferogram thus encompasses a series of RF spectral combs spectrally isolated from each other for simultaneous broadband high-resolution ringdown measurements.

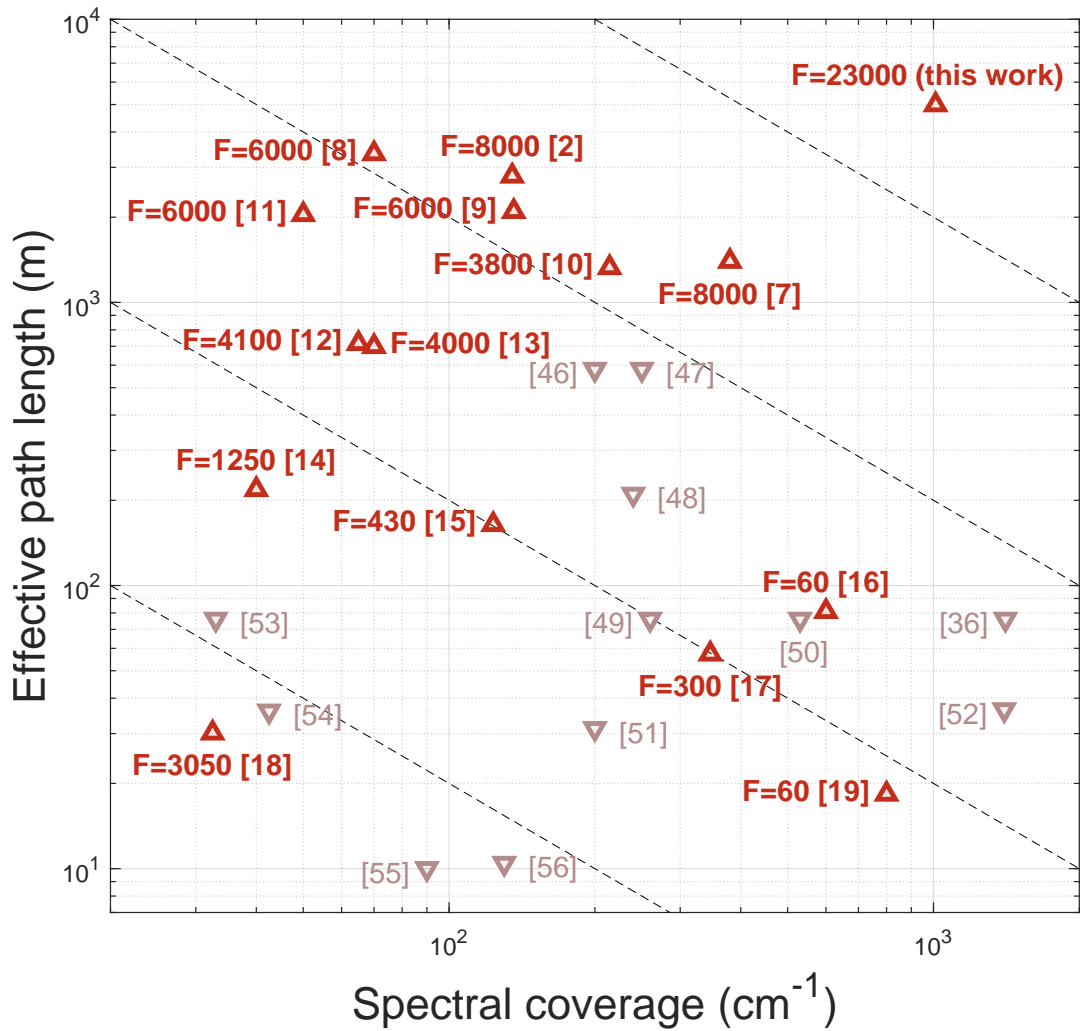

**Figure 2: Multi-species trace gas sensing at the extreme performance level**. A larger spectral coverage in the mid-infrared region (2-20 µm) enables more molecular species to be identified, while a larger effective absorption path length enables molecules of lower concentrations to be detected. Summarized[2,7-19,36,46-56] are apparatus demonstrated with spectroscopy results performed over 10 cm$^{-1}$ coverage and over 10 meters effective path length. Only tabletop mid-infrared frequency comb spectroscopy experiments are considered. For all plotted data points, experiments performed with enhancement cavities (or multi-pass cells) are labelled with up-pointing (or down-pointing) triangles in bright red (or dark red) color. Maximum cavity finesse is indicated for experiments using enhancement cavities. The black dashed lines indicate trends at constant product between effective path length and spectral coverage.

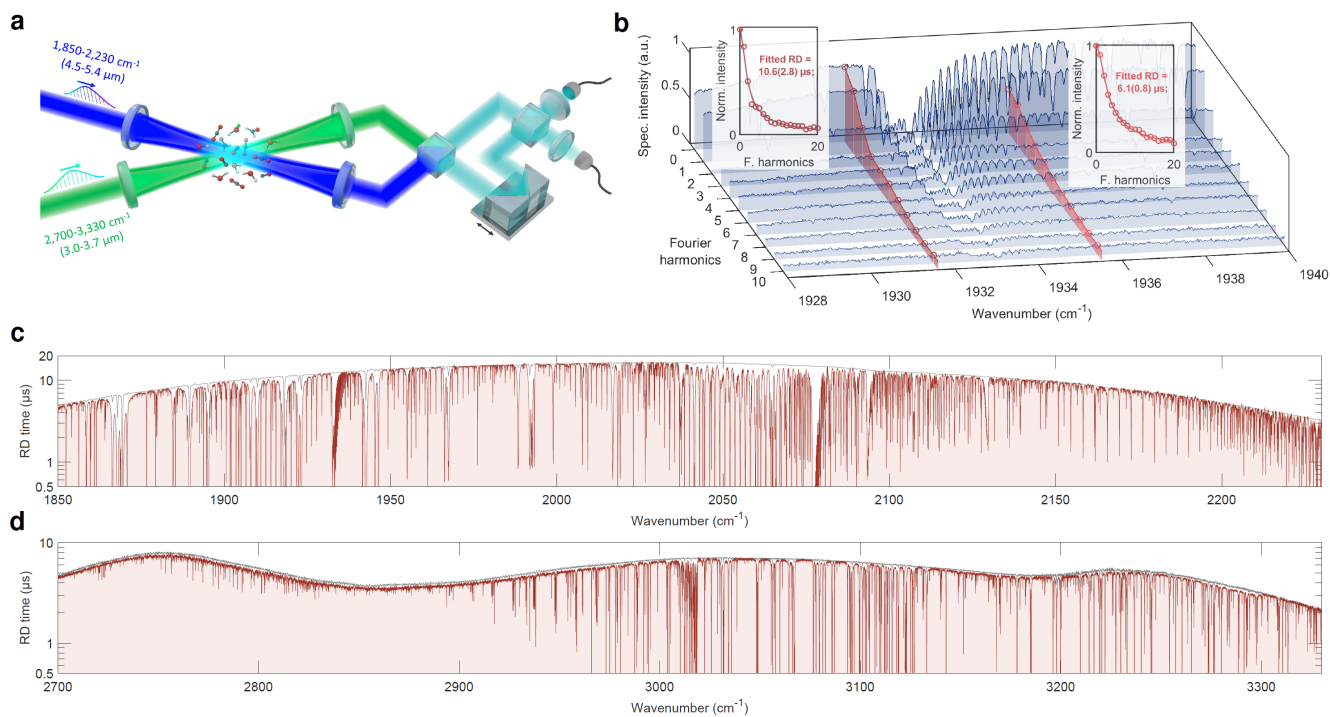

**Figure 3: Apparatus and survey ringdown spectra collected with the MRCI technique. a,** Apparatus: two sets of high-finesse cavities are used for probing the same gas sample. One at 5-μm wavelength range (finesse ~23,000) permits spectral output between 1,850-2,230 cm$^{-1}$, the other at 3-μm wavelength range (finesse ~14,000) permits 2,700-3,330 cm$^{-1}$. **b,** Sample spectral data processed from the recorded interferogram data collected at 2 seconds acquisition time, 800 MHz instrument resolution, with the cavities loaded with a breath sample. Trend lines in red highlight Fourier components from different cavity resonances. The trend of decay in spectral intensities with the increase of Fourier harmonics order are fitted to an analytically derived formula to determine the ringdown times. Insets show the fitted ringdown times, where circles are experimental data, solid lines are the fitted curves. **c** and **d,** Survey ringdown spectra measured for the same breath samples (red) employing the two sets of cavities. Data in grey shows the ringdown data measured for cavity held at base pressure (below 3 mTorr). The decrease in ringdown times in data measured for breath from that measured for cavity at base pressure gives molecular absorption spectrum measured from the loaded breath sample.

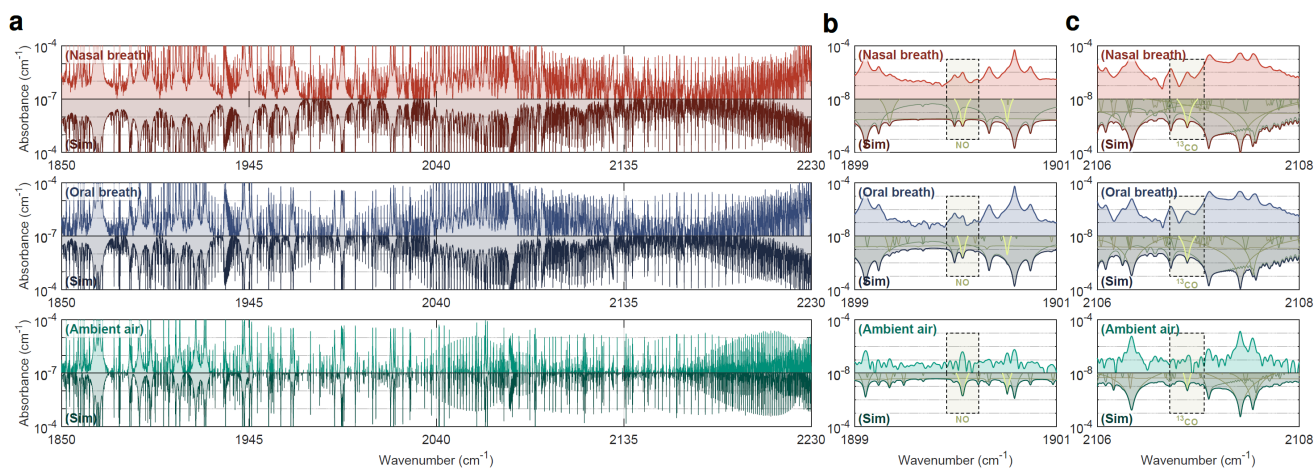

**Figure 4: Molecular spectroscopy from the 5-µm cavity.** Shown are Nasal breath (red; first row), oral breath (blue; second row), and ambient air (green; third row). **a,** Spectroscopy data from 1,850 to 2,230 cm$^{-1}$. Experimental data is compared to the global fit results plotted inverted in signs at darkened colors. **b** and **c,** Zoom-ins to individual spectral ranges. Fit results from individual molecules are plotted on top of the global fit results. Color highlight in bright yellow is given to absorption features from nitric oxide (NO) in **b** and carbon monoxide isotopologue ($^{13}$CO) in **c**.

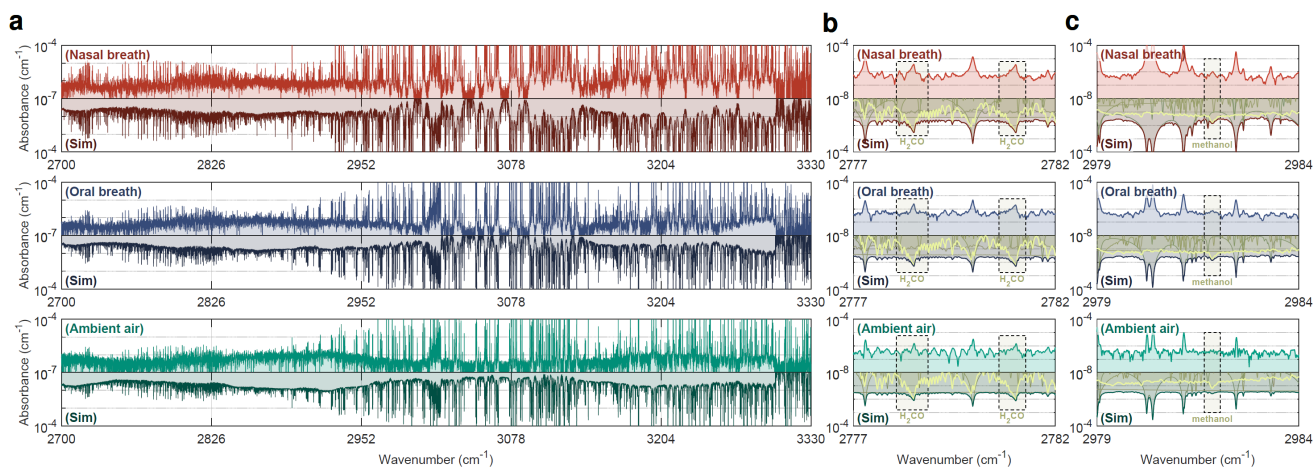

**Figure 5: Molecular spectroscopy from the 3-µm cavity.** Shown are nasal breath (red; first row), oral breath (blue; second row), and ambient air (green; third row). **a,** Spectroscopy data from 2,700 to 3,330 cm$^{-1}$. Experimental data is compared to the global fit results plotted inverted in signs at darkened colors. **b** and **c,** Zoom-ins to individual spectral ranges. Fit results from individual molecules are plotted on top of the global fit results. Color highlight in bright yellow is given to absorption features from formaldehyde (H$_2$CO$_3$) in **b** and methanol (CH$_3$OH) in **c**.

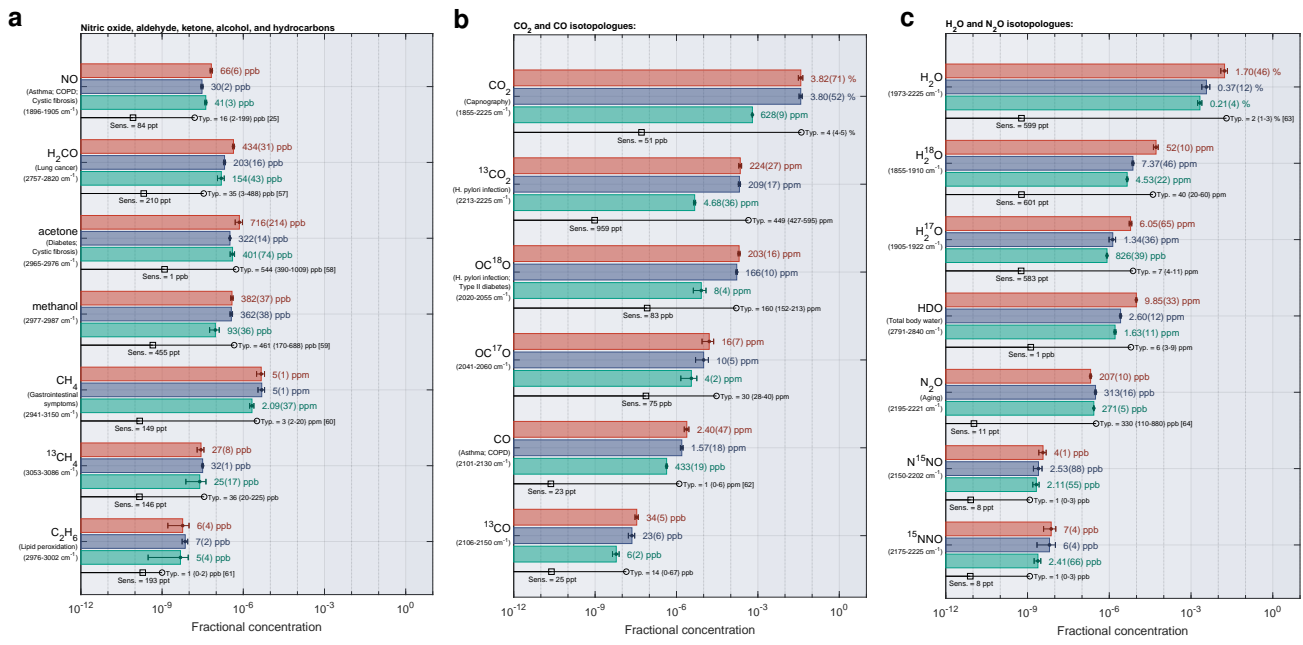

**Figure 6: Molecular concentrations summary totaling 20 species for nasal breath (red bars), oral breath (blue bars), and ambient air (green bars).** Left, middle, right panels summarize different molecular species. For each species, the fitted concentrations from our comb-based spectroscopy scans and their known clinical relevance for breath analysis are given. On the stem lines, typical concentrations[25,57-64] found in exhaled breath are indicated with open circles (the numbers inside the curly brackets provide the range of concentrations), while experimental detection sensitivity limits are indicated with open squares.

## Methods

**Modulated Ringdown Comb Interferometry (MRCI).** *Theoretical model.* Consider a high-finesse cavity made of two identical cavity mirrors is coupled with a passively-stable frequency comb. The cavity length is modulated by a sinusoidal wave at rate $\omega_m$ centered at the cavity length of $d_0$. The mean cavity free spectral range is matched to the comb spacing. The cavity resonance frequency modulation depth is smaller than the comb spacing but is large enough to permit the instantaneous comb bandwidth to resonant couple through the cavity twice per modulation period: once when the cavity length is swept up, the other down. For simplicity, imagine for now the incident comb is shut off during the cavity length down sweep (more discussions later). Transmission bursts from all cavity resonances thus have the same time periodicity $T = 2\pi/\omega_m$ and emitted all from the cavity up sweep. The electric field intensity components $|E_{cav,i}(t)|^2$ in time $t$ from the $i$-th cavity resonance can be decomposed into a Fourier series:

$$|E_{cav,i}(t)|^2 = \sum_{n=-\infty}^{+\infty} f_{ni} \exp(jn\omega_m t), \qquad (1)$$

where $f_{ni}$ is the $n$-th Fourier component from the $i$-th resonance, and $j$ denotes imaginary unit.

The Michelson interferometer introduces an additional phase modulation specific to the optical frequency. The laser intensity vs. time readout by the photodetectors in the Michelson interferometer is expressed as $\sum_i |E_{cav,i}(t)|^2 [1 \pm \cos(\omega_i t)]$, where $\omega_i = k_i V$ is the Doppler frequency, $k_i$ is the optical wavenumber, and $V$ is the optical path length difference scanning speed assumed to be a constant (more discussions later). The signal carriers are spectrally detected at RF frequencies over $|n\omega_m \pm \omega_i|$ and the non-cosine modulated terms at $|n|\omega_m$. They are spectrally isolated if $\omega_i$ for all $i$ are non-equal to integer multiples of $\omega_m/2$. Spectral intensities measured by the non-cosine modulated terms can be used to determine detectors' relative response function to implement digital-autobalancing. The balanced intensity output $I(t)$ after cancelling out the non-cosine modulated terms is given by:

$$I(t) = 2 \sum_i \sum_{n=-\infty}^{+\infty} f_{ni} \exp(jn\omega_m t) \cos(\omega_i t). \qquad (2)$$

The expression of $|E_{cav,i}(t)|^2$ for a single transmission burst (i.e., cavity resonance swept through a comb line for once) is previously derived in Ref.[65]. Omitting common proportionality and assuming $T \gg \tau_{RD,i}$, $f_{ni}$ can be calculated from:

$$f_{ni} = \int_{-\infty}^{\infty} \exp\left(-\frac{t'}{\tau_{RD,i}}\right) |\text{erfc}[\Lambda_i(t')]|^2 \exp(-jn\omega_m t')dt', \tag{3}$$

where $\tau_{RD,i} = \tau/(-2\ln\rho_i)$ is the cavity ringdown time, $\tau = 2d_0/c$ the cavity round trip time, $c$ the speed of light, $\rho_i = r_i^2 \exp[-(\alpha_i/2) \cdot 2d_0]$ the round-trip field loss, $r_i$ the mirror reflection coefficient, $\alpha_i$ the intracavity absorption coefficient, and erfc the complementary error function. The $\Lambda_i(t')$ is given by

$$\Lambda_i(t') = \frac{1-j}{2\sqrt{2}}\left(\frac{1}{k_i v_i \tau}\right)^{1/2} \cdot (-\ln\rho_i) - \frac{1+j}{2\sqrt{2}}\left(\frac{1}{k_i v_i \tau}\right)^{1/2} \cdot 2k_i v_i t', \tag{4}$$

where $v_i$ is the cavity length swept velocity when the $i$-th resonance is scanned to the comb line.

Equation (3) is used to fit the experimental data for $|f_{ni}|$ at the same $i$ but different $n$ to find $\tau_{RD,i}$. For molecular spectroscopy with the goal to determine $\alpha_i$, one compares the ringdown times measured with the cavity loaded ($\alpha_i \neq 0$) and unloaded ($\alpha_i = 0$) with a gas sample. From dependence of $\tau_{RD,i}$ on $\rho_i$ one finds $\alpha_i$ from:

$$\alpha_i = \frac{1}{c}\left[\frac{1}{\tau_{RD,i}(\alpha_i \neq 0)} - \frac{1}{\tau_{RD,i}(\alpha_i = 0)}\right]. \tag{5}$$

*Experimental implementation.* The MRCI technique was implemented in our experiment with the incident comb power passively stayed on, a high-finesse cavity swept locked to the incident comb lines, and a passively-scanning Michelson interferometer. The $\omega_m$ and $\omega_i$ are not perfectly coherent. Below we detail their sources of incoherence and present our data analysis strategies. These strategies are the key to remove the experimental needs to ensure perfect coherence in $\omega_m$ and $\omega_i$ that can require dramatically increased experimental complexity and decrease robustness.

1) The $\omega_m$: Due to the presence of intracavity dispersion and piezo hysteresis, comb lines on cavity resonance at an earlier time during the cavity length up sweep are on resonance at a later time during the cavity down sweep. While two transmission bursts are emitted from each resonance per cavity length round trip modulation period $T$, intensity component $|E_{cav,i}(t)|^2$ generally repeats at periodicity $T$ rather than $T/2$: the base pattern in repetition is that from a cavity up and subsequent down sweeps. For equation (3) to be directly applicable, we process the transmission bursts from the up sweep separately from the down sweep. A logical TTL function phase-synchronized with the cavity modulation signals is digitally generated in the data post-processing to pick up transmission bursts solely from the up (or down) sweep via multiplication to the raw interferogram data. This is equivalent to having the incident comb source shut off during the down (or up) sweep. The two sets of interferogram, one containing only the up-sweep bursts and the other only the down-sweep bursts, are processed separately by equation (3). On a separate note, when it's of interest to implement dispersion spectroscopy[66], one can process the interferogram without separating the up and down bursts. Equation (3) would need to be modified to account for an emission time delay between adjacent up and down bursts from the expected time separation at $T/2$. This quantity measures intracavity dispersion.

2) The $\omega_i$: Conventional delay stage introduces considerable mechanical jitters causing $\omega_i$ to be fluctuating in time. The jittering of $\omega_i$ is small compared with $\omega_m$ but must be properly accounted in the data processing for applications at high spectral resolution. Equation (2) should be modified into a more general form by replacing $\omega_i t$ with $k_i z$:

$$I(t,z) = 2 \sum_i \sum_{n=-\infty}^{+\infty} f_{ni} \exp(jn\omega_m t) \cos(k_i z) \qquad (6)$$

Here $z$ is the optical path length difference scanned. In the experiment, the real time mapping from $z$ to $t$ can be measured with a wavelength-stable continuous-wave (CW) laser co-propagating with the comb light into the Michelson interferometer. With both interferogram and space variable $z$ sampled by the same time array, explicit dependence of interferogram on time can be removed using demodulation: first multiply the interferogram $I(t,z)$ with digitally-generated $\exp(-jn'\omega_m t)$ sampled over the same time array, then bandpass centered at the Doppler frequencies $\{\omega_i\}$. This results in,

$$\{I(t,z) \cdot \exp(-jn'\omega_m t)\}_{\text{bandpass}} = \sum_i f_{n'i} \cos(k_i z). \qquad (7)$$

The demodulated output is a complex function of single variable $z$ and can be non-uniformly Fourier transformed to find the spectrum of $|f_{n'i}|$ vs. $k_i$. The value of $|f_{n'i}|$ for all $i$ can thus be determined. The demodulation is repeated at different $n'$ spanning from negative to positive integers: The $|f_{n'i}|$ and $|f_{-n'i}|$ are independently measured by two signal carriers at $n'\omega_m + \omega_i$ and $n'\omega_m - \omega_i$ but probe the same physical quantity $|f_{n'i}| = |f_{-n'i}|$ (see equation (3)). They can be averaged together to improve the signal-to-noise ratio.

To summarize: the raw interferogram is first linearly decomposed into two interferograms: one contains bursts exclusively from the up sweep, the other down. The two interferograms are respectively processed to find $|f_{ni}|$ containing non-negative $n$ using the demodulation technique. The two sets of $|f_{ni}|$ are averaged together and equation (3) is applied to find ringdown time $\tau_{RD,i}$.

**Apparatus.** *Frequency combs.* The comb used for the data collection in the 2,700–3,330 cm$^{-1}$ spectral range is a singly-resonant optical parametric oscillator (OPO) built from a periodically poled lithium niobate crystal and synchronously pumped by an Ytterbium comb at 137 MHz repetition rate centered at 1064 nm and pulse duration of 100 fs. This OPO is previously reported in Ref.[24]. The instantaneous bandwidth is 180 cm$^{-1}$. Spectral tuning is performed by translating the crystal location. The comb used for the 1,850–2,230 cm$^{-1}$ spectral range is built from a new singly-resonant OPO using a Zinc Germanium Phosphide (ZGP) crystal synchronously pumped by a Thulium comb at 110 MHz repetition rate centered at 1,960 nm and pulse duration of 1,000 fs. This comb source is developed because oxides materials are incapable of generating sufficient power per comb tooth at ≥ 4.8 µm wavelength[17,67]. The ZGP OPO is designed following the same strategy previously documented in Ref.[68]. Both the spatial walk-off from birefringence and the temporal walk-off from group velocity mismatch are taken into account to design the threshold pump power sufficiently small compared to the maximum available pump power. Ring cavity geometry rather than linear is adopted to avoid round-trip signal absorption loss from one additional pass through the crystal. The cavity is unpurged but sealed for passive frequency stability. The OPO had an instantaneous bandwidth of 60 cm$^{-1}$ and is broadly tunable in the range of 1,850–2,230 cm$^{-1}$ simply by translating the OPO cavity length to manipulate the group delay dispersion. Limited power below 1,850 cm$^{-1}$ is due to water absorption and above 2,230 cm$^{-1}$ due to carbon dioxide absorption. When idler is tuned to 2,040 cm$^{-1}$ (4.9 µm), power per comb tooth up to 13 µW is measured. This is more than a factor of ten higher than all comb sources ever reported near the 5 μm wavelength[36,69-73]. For additional details about the OPO, see Extended Data Fig. 1.

*Data acquisition.* High-reflectivity mirrors for both cavities were newly purchased from Lohnstar Optics (technical communications only; not endorsement). Both cavities have their mean free spectral range matched to twice the frequency spacing of the incident combs. The 5-μm cavity is 68 cm in length and modulated at 13 kHz. The 3-μm cavity is 55 cm in length and modulated at 18 kHz. The modulation rates are set sufficiently low to enable sufficient time for each ringdown event to eventually decay down to the detector noise floor. Modulation rate is higher for the 3-μm cavity due to smaller cavity ringdown times overall. Servo error signals generated from the photodetectors in the Michelson interferometer are summed and demodulated at the third harmonics of the modulation frequencies for feedback stabilization of the cavity lengths to the comb line frequencies. Both combs are free-running. The OPO spectral center and the interferometer's optical path length difference (OPD) scanning speed are feedforward adjusted automatically. The feedforward map ensures every interferogram scan is executed such that the Doppler frequencies stayed about the same and are sufficiently detuned from integer multiples of half the cavity length modulation rates to avoid spectral overlapping issues. For each scan the OPD scanning speed is no faster than 200 mm/s. A total of three data channels low passed at 1 MHz are collected simultaneously at 2 MSPS sampling rate and 16-bit resolution: interference fringes for the comb light using two photodetectors, and for the CW laser with one photodetector. No photodetectors require cooling by liquid nitrogen. Data collection for the entire 5-μm spectral range (1,850-2,230 $cm^{-1}$) is fully automated with servo loops robustly sustained throughout. For the 3-μm range (2,700-3,330 $cm^{-1}$), however, the picomotor actuator used for adjusting the crystal location introduces considerable mechanical jitter that occasionally disengage the servo. We expect future upgrades to crystal positioning and/or implementing automatic servo engagement can fully automate data collection also for the 3-μm region. Breath and air data are processed at 800 MHz instrument resolution and acquisition time per interferogram is about 2 seconds. Full survey spectrum for the 5-μm spectral range at 380 $cm^{-1}$ coverage is collected with ~500 interferograms, while ~1,400 interferograms for the 3-μm range at 630 $cm^{-1}$ coverage. Empty cavity survey data is collected at 2 GHz instrument resolution. For all experimental data the demodulation is performed for the 5-μm data up to the 20-th Fourier harmonics and the 3-μm data up to the 15-th harmonics. This utilized signal carriers measured with high signal-to-noise ratio up to 300 kHz electronic bandwidth. Fourier harmonics spectra collected at different spectroscopic regions are summed before determining the full coverage ringdown spectrum. Data analysis is performed using a 32-core 3975WX CPU and a RTX A6000 GPU. The GPU is used for accelerating demodulation and Fourier transforms. The multi-core CPU is used for accelerating Ringdown extraction. An example interferogram data is presented in Extended Data Fig. 2.

**Sample collections and handling.** The study was approved by the Institutional Review Board of University of Colorado Boulder with protocol number 23-0536. Potential subjects must be above 18 years old to participate. No exclusion criteria. A research subject was recruited from flyers. After written informed consent was obtained, the participant was contacted by a research member via emails for the scheduling of appointment time for breath samples collection. Right before sample collection, participants filled in an online questionnaire for the collection of demographics information and basic medical history. All information collected from the subject was stored and managed electronically via RedCap[74,75], a secure server for human research data management. All data exported from RedCap is de-identified. A study-ID is used for correlating the questionnaire information with the breath analysis measurement results. The research participant was advised to refrain from consumption of food or drink (water excluded) and usage of mouthwash or cough drops one hour before providing breath samples. Two breath samples were collected: one by inhaling and exhaling through the nose but with the mouth closed (i.e., nasal breath sample); the other by inhaling and exhaling through the mouth but with the nose closed (i.e., oral breath sample). The two samples were collected immediately adjacent to each other, both by inhaling to full lung capacity and followed by exhaling only the end tidal breath into a Tedlar bag. The ambient air sample was collected on a separate day and also using a Tedlar bag. All collected gas samples were analyzed immediately within the day of their collections. Used Tedlar bags were securely autoclaved and discarded. The chamber and cavity mirrors were flushed with ultra-high purity argon gas at 4 liters per minute for 10 minutes after each sample analysis to prevent potential cross-sample contamination. No degradation is observed for the cavity finesse.

**Molecular line fitting.** The HITRAN database (edition 2020) is used to extract concentrations totaling 20 molecular species. Methanol (214.2 K, 102.7 Torr) and acetone (209.8 K, 109.6 Torr) use cross sectional data directly measured at specific temperature and pressure due to lack of line intensity data. For the other 18 species, cross sectional data is calculated from the line intensity data using Voigt profile evaluated at our experimental conditions (293.15 K, 100 Torr). Pressure broadening by air is considered, while self-pressure broadening and pressure-shift of the line frequency centers were ignored. Molecular line fitting is performed bearing in mind that:

i) We may not be fitting all species that are detected, which could partly be attributed to lack of cross-sectional data. For exhaled breath, more than 1,000 species have been reported and over 40 % are hydrocarbons that can be spectroscopically detected in the 3 to 4 μm wavelength range[20,76]. Unfitted species that are potentially present in our experimental data are generally large molecules exhibiting state-unresolved ultra-broadband absorption features.

ii) Strongly-absorbing species such as water and carbon dioxide produced saturated absorption features. Errors in the prediction of absorption lineshape at far-off line centers[26] prohibits strongly absorbing species to have their concentrations determined over their saturated absorption regions. Further, other species that are weakly-absorbing could also be impacted when their line centers spectrally overlap with the saturated absorption features.

We have thus developed the following fitting strategy: We implemented a sliding window to select out 10 $cm^{-1}$ spectral range at a time for molecular line fitting. The window is step incremented at 0.5 $cm^{-1}$ to uniformly sample the whole coverage (1,010 $cm^{-1}$). Least squares fitting is performed with residuals given by the difference in the logarithm of absorption spectra. After fitting all windows separately, a slowly-varying baseline is introduced for the whole coverage to vertically offset the simulated data to better match the experimental data. Here, the baseline is to mimic the absorption features from unknown species and correct mis-predicted absorption lineshape from the fitted species, both assumed to have a bandwidth broader than the window width at 10 $cm^{-1}$. With its introduction the fittable species can be determined at minimized influence from these broadband features. We re-fit the spectrum with the simulated data added to the baseline in each window to compare it to the experimental data. After all windows are fitted the baseline is updated again. The whole process is iterated five times for convergence. Fitted concentrations from the final iteration run are used for determining molecular concentrations. For each species, its concentration is determined from windows where it can be reliably fitted. For the two saturated species $CO_2$ and $H_2O$, windows where their max absorption coefficient exceeds $10^{-4}$ $cm^{-1}$ are discarded from statistical evaluations. For the other 18 species, only windows where their max absorption cross sections are within 70 % of their strongest absorption cross section in the whole coverage are considered. The cutoff percentage is chosen sufficiently low so that each species has its concentration evaluated generally from more than 30 windows for statistical convergence. For each window the averaging weight is given by its max absorption cross section. The selected window ranges for each of the 20 species are reported in Figure 6 in the main text.

**References:**

65  Poirson, J., Bretenaker, F., Vallet, M. & LeFloch, A. Analytical and experimental study of ringing effects in a Fabry-Perot cavity. Application to the measurement of high finesses. *J Opt Soc Am B* **14**, 2811-2817 (1997). doi:10.1364/Josab.14.002811

66  Wysocki, G. & Weidmann, D. Molecular dispersion spectroscopy for chemical sensing using chirped mid-infrared quantum cascade laser. *Opt Express* **18**, 26123-26140 (2010). doi:10.1364/Oe.18.026123

67  Schunemann, P. G., Zawilski, K. T., Pomeranz, L. A., Creeden, D. J. & Budni, P. A. Advances in nonlinear optical crystals for mid-infrared coherent sources. *J Opt Soc Am B* **33**, D36-D43 (2016). doi:10.1364/Josab.33.000d36

68  Iwakuni, K. *et al.* Phase-stabilized 100 mW frequency comb near 10 μm. *Appl Phys B* **124** (2018). doi:10.1007/s00340-018-6996-8

69  Vasilyev, S. *et al.* Longwave infrared (6.6-11.4 μm) dual-comb spectroscopy with 240,000 comb-mode-resolved data points at video rate. *Opt Lett* **48**, 2273-2276 (2023). doi:10.1364/Ol.477346

70  Smolski, V. *et al.* Half-Watt average power femtosecond source spanning 3-8 Aμm based on subharmonic generation in GaAs. *Appl Phys B* **124** (2018). doi:10.1007/s00340-018-6963-4

71  Heckl, O. H. *et al.* Three-photon absorption in optical parametric oscillators based on OP-GaAs. *Opt Lett* **41**, 5405-5408 (2016). doi:10.1364/Ol.41.005405

72  Leindecker, N. *et al.* Octave-spanning ultrafast OPO with 2.6-6.1μm instantaneous bandwidth pumped by femtosecond Tm-fiber laser. *Opt Express* **20**, 7046-7053 (2012). doi:10.1364/Oe.20.007046

73  Vodopyanov, K. L., Sorokin, E., Sorokina, I. T. & Schunemann, P. G. Mid-IR frequency comb source spanning 4.4-5.4 μm based on subharmonic GaAs optical parametric oscillator. *Opt Lett* **36**, 2275-2277 (2011). doi:10.1364/Ol.36.002275

74  Harris, P. A. *et al.* Research electronic data capture (REDCap)-A metadata-driven methodology and workflow process for providing translational research informatics support. *J Biomed Inform* **42**, 377-381 (2009). doi:10.1016/j.jbi.2008.08.010

75  Harris, P. A. *et al.* The REDCap consortium: Building an international community of software platform partners. *J Biomed Inform* **95** (2019). doi:10.1016/j.jbi.2019.103208

76  Drabinska, N. *et al.* A literature survey of all volatiles from healthy human breath and bodily fluids: the human volatilome. *Journal of Breath Research* **15** (2021). doi:10.1088/1752-7163/abf1d0

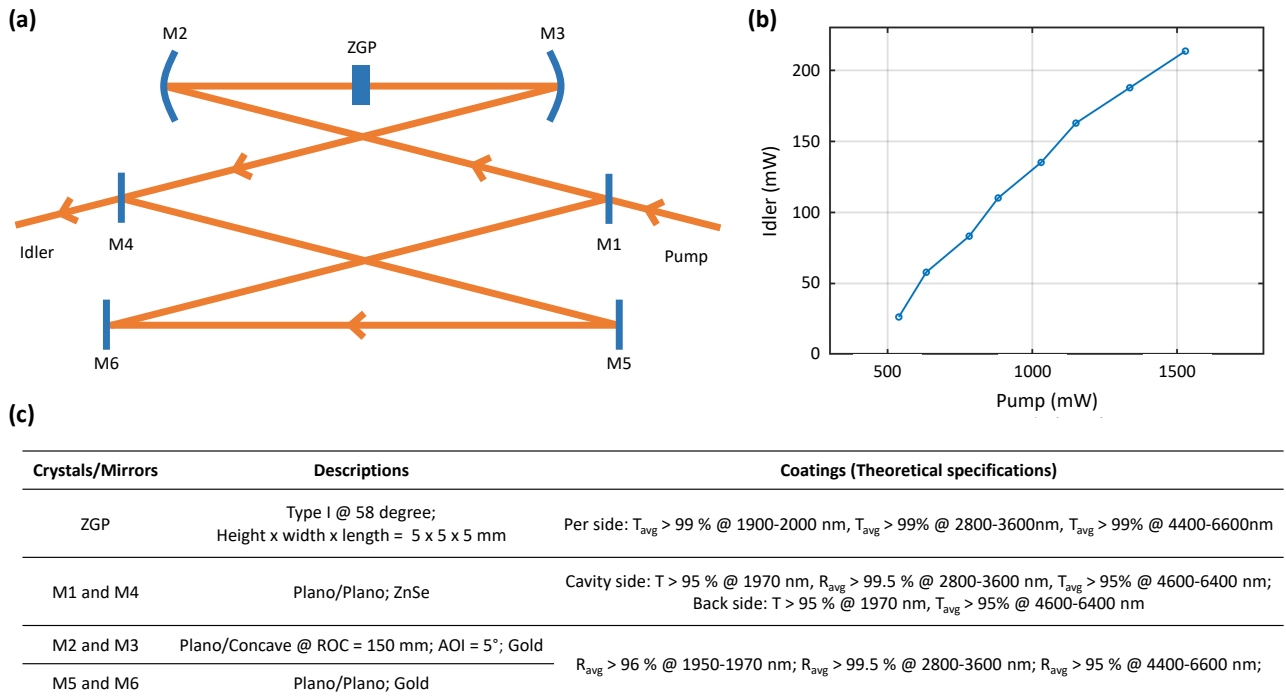

**Extended Data Fig. 1: The Tm-ZGP OPO comb. a,** The OPO cavity geometry. Cavity mirrors are labeled from M1 to M6. Pump light is injected from M1 and idler light is out-coupled from M4. Coarse and fine cavity length control are achieved with a picomotor and a piezo respectively mounted on M5 and M6. **b,** Power dependence measured for the idler on the pump, when the idler is tuned to 2,040 cm$^{-1}$. The highest idler average power of 214 mW is measured at the pump power of 1,530 mW. The threshold pump power is 340(50) mW. Slope power efficiency is 19(1) % and photon conversion efficiency is 49(2) %. **c,** Table summary of the crystals and mirrors specifications.

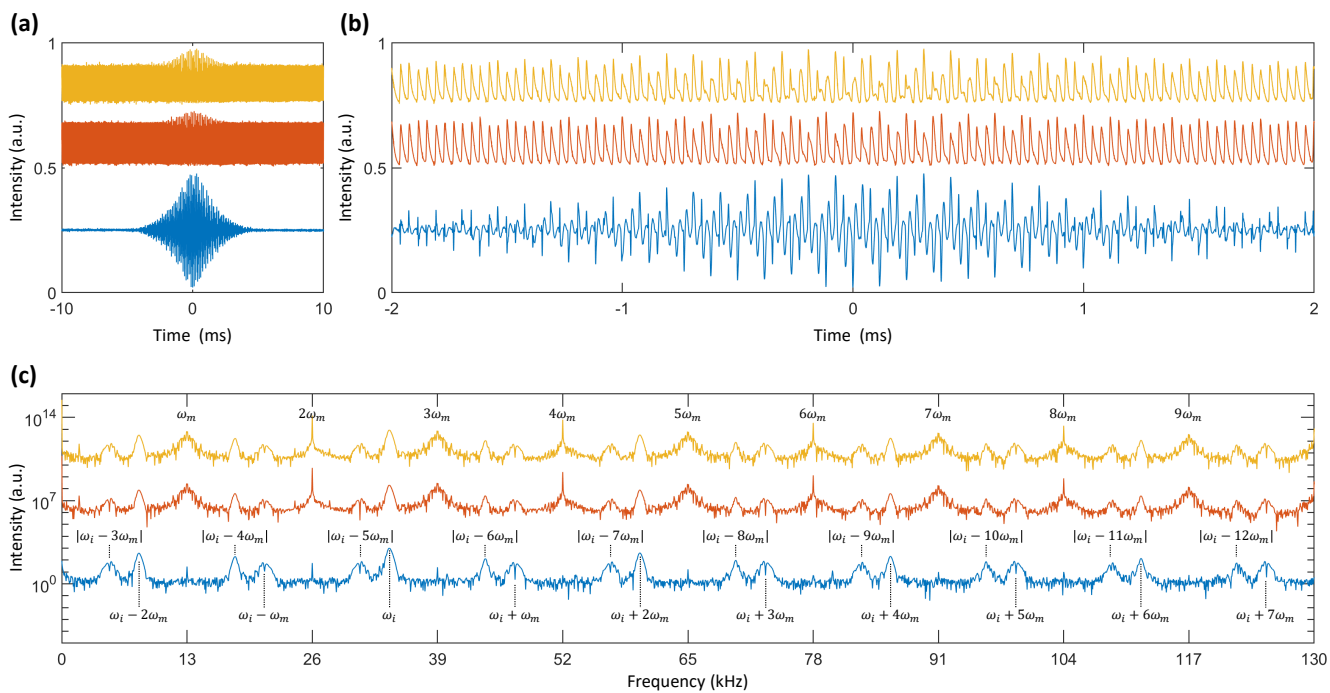

**Extended Data Fig. 2: Interferogram data. a,** Experimental interferogram raw data collected as a function of time for the two photodetectors measuring the cavity transmitted comb light (yellow and orange). Blue trace is the balanced output obtained from data post-processing. **b,** Zoomed in of (a) into ±2 ms. **c,** Fourier-transformed spectra for data in **a** over the time span of 20 ms. Signal carriers at $n\omega_m \pm \omega_i$ and non-cosine modulated terms at $n\omega_m$ are labeled. For this data $\omega_m$ = 13 kHz and $\omega_i$ ~ 34 kHz. All traces in **a** - **c** are vertically offset for clarity.